\documentclass[final,1p,times]{elsarticle} 

\usepackage{graphicx}
\usepackage{amssymb} 
\usepackage{amsthm} 
\usepackage{lineno}
\usepackage{ulem}
\usepackage{color}
\usepackage[usenames,dvipsnames,svgnames,table]{xcolor}

\journal{Nuclear Physics A} 

\begin{document}

\begin{frontmatter} 

\title{Baryon number and charge fluctuations from lattice QCD}

\author{Christian Schmidt (for the BNL-Bielefeld\fnref{col1} Collaboration)}
\fntext[col1] {Collaboration members are: A. Bazavov, H.-T. Ding, P. Hegde, O. Kaczmarek, F. Karsch, E. Laermann, 
S. Mukherjee, P. Petreczky, C. Schmidt, D. Smith, W. Soeldner,  M. Wagner.}
\address{Universit\"at Bielefeld, Fakult\"at f\"ur Physik, Postfach 100131, D-33501 Bielefeld, Germany}

\begin{abstract} 
We calculate electric and baryonic charge fluctuations on the lattice. Results have been obtained with the highly improved 
staggered quark action (HISQ) and almost physical quark masses on lattices with spacial extent of $N_\tau=6,8,12$.
Higher order cumulants of the net-charge distributions are increasingly dominated by a universal scaling behavior, which is 
arising due to a critical point of QCD in the chiral limit.  Considering cumulants up to the sixth order, we observe that they  
generically behave as expected from universal scaling laws, which is quite different from the cumulants calculated within the 
hadron resonance gas model. Taking ratios of these cumulants, we obtain volume independent
results that can be compared to the experimental measurements. Such a comparison will unambiguously relate the QCD transition
temperature that has been determined on the lattice with the freeze out temperature of heavy ion collision at LHC and RHIC. 
\end{abstract} 

\end{frontmatter} 


\section{Introduction}
\begin{figure}[htbp]
\begin{center}
\begin{minipage}{0.4\textwidth}
\begin{center}
\includegraphics[width=0.99\textwidth]{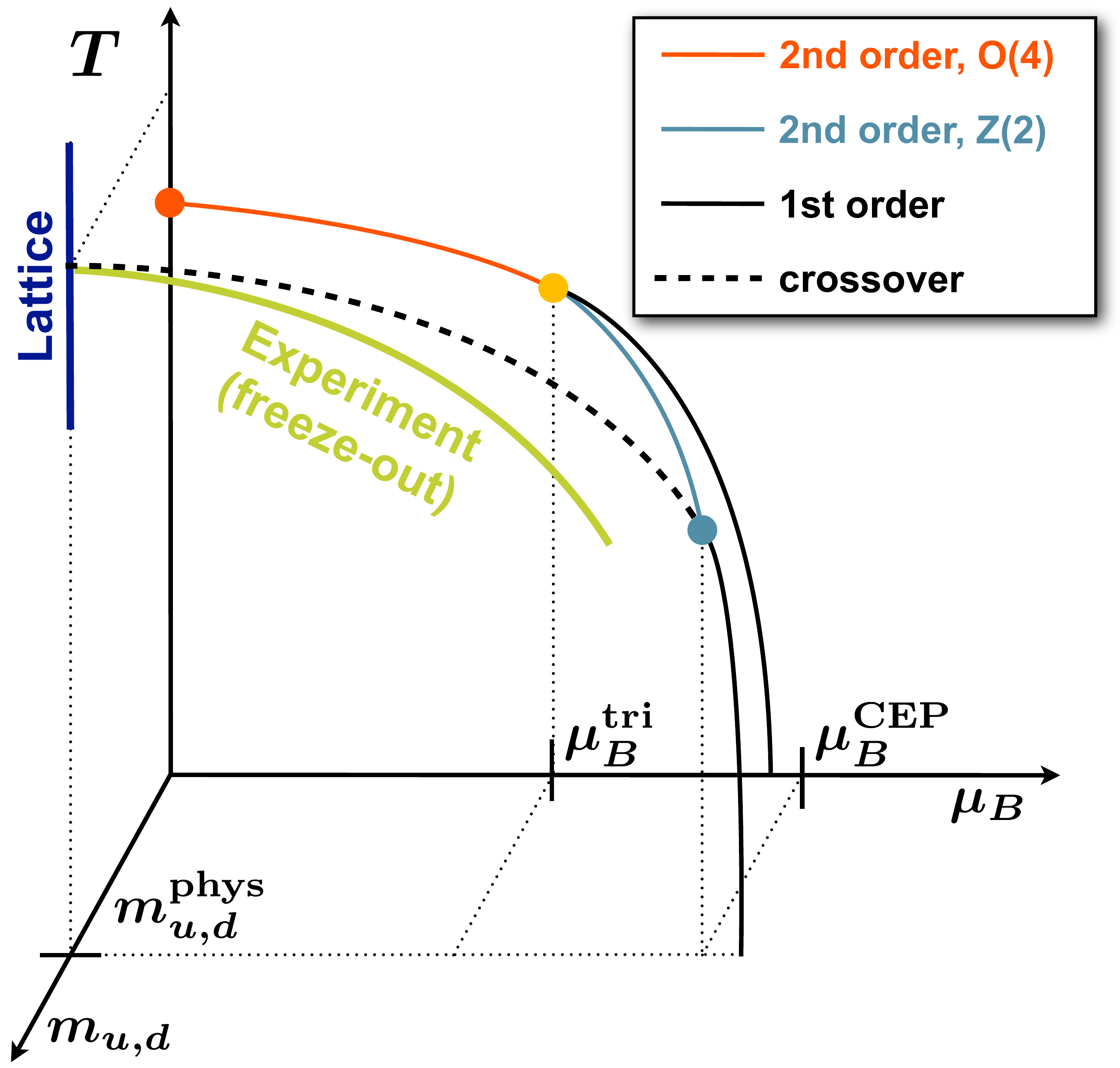}
\end{center}
\end{minipage}
\end{center}
\caption{Generic phase diagram of QCD, based on model calculations and model independent symmetry arguments.  Also indicated are the regions in the phase diagram where we are able to obtain results on fluctuation observables from lattice QCD and experiments, respectively. \label{fig:pdiag}}
\end{figure}

Calculating the phase diagram of strongly interacting matter is one of the most important and outstanding problem of non-perturbative QCD. A generic plot of the QCD phase diagram based on model calculations and model independent symmetry arguments is shown in Fig.\ref{fig:pdiag}.
Due to the notorious sign problem lattice QCD calculations are currently restricted to zero baryon chemical potential. However, unlike in heavy ion experiments, one can vary the quark mass in lattice QCD simulations. As one expects a critical point in the phase diagram at nonzero density, fluctuation observables are the most natural choice in order to detect QCD critical behavior. Their divergent behavior can give insight in the structure of the phase diagram. Currently various experimental programs aim at probing the phase diagram at nonzero baryon number density, such as the low energy scan at RHIC, NA61 at the SPS as well as future experiments at FAIR and NICA. In experiments one has access to the fluctuations of conserved charges at freeze out. The freeze-out curve in the phase diagram can be parametrized as a function of collision energy \cite{fc}. 

\section{Cumulants of conserved charges}
\begin{figure}[htbp]
\begin{center}
\begin{minipage}{0.294\textwidth}
\begin{center}
\includegraphics[width=0.998\textwidth]{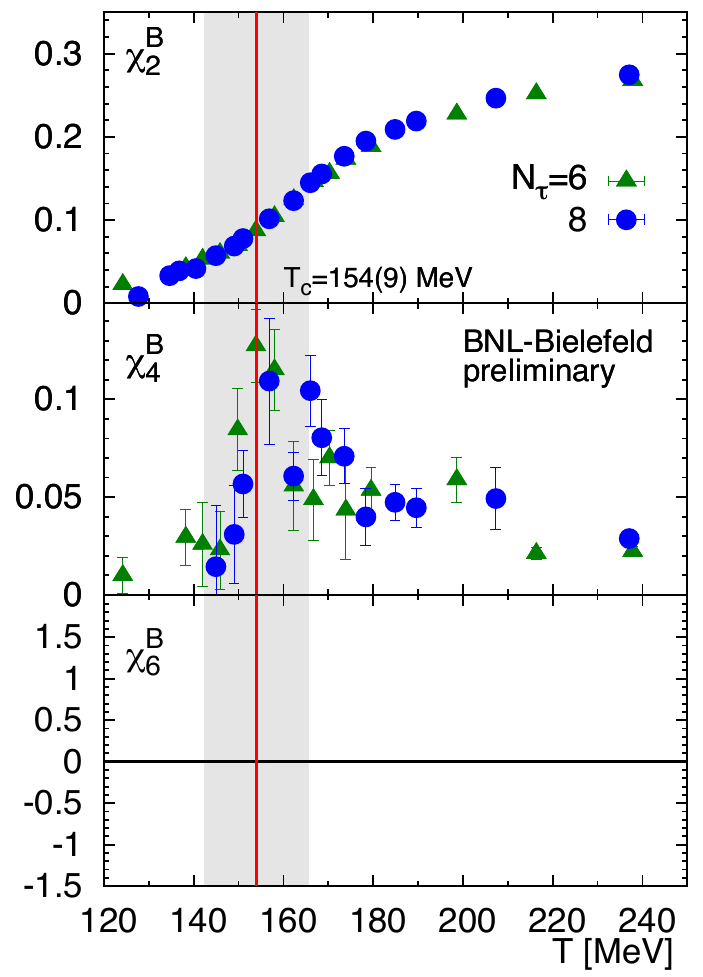}
\end{center}
\end{minipage}
\begin{minipage}{0.294\textwidth}
\begin{center}
\includegraphics[width=0.998\textwidth]{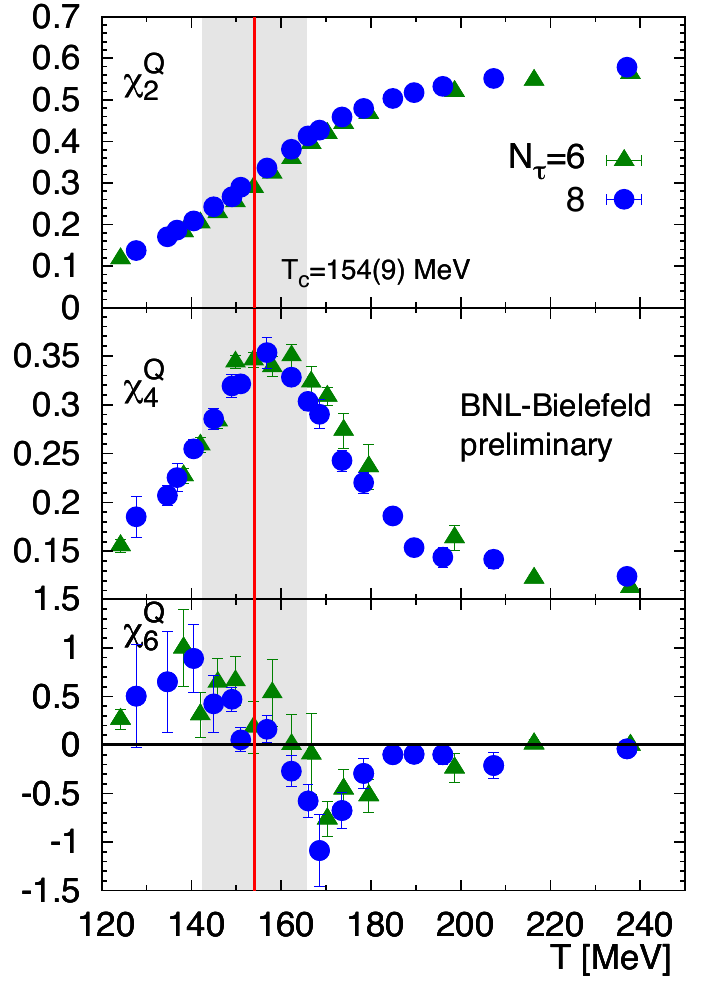}
\end{center}
\end{minipage}
\begin{minipage}{0.4\textwidth}
\begin{center}
\includegraphics[width=0.998\textwidth]{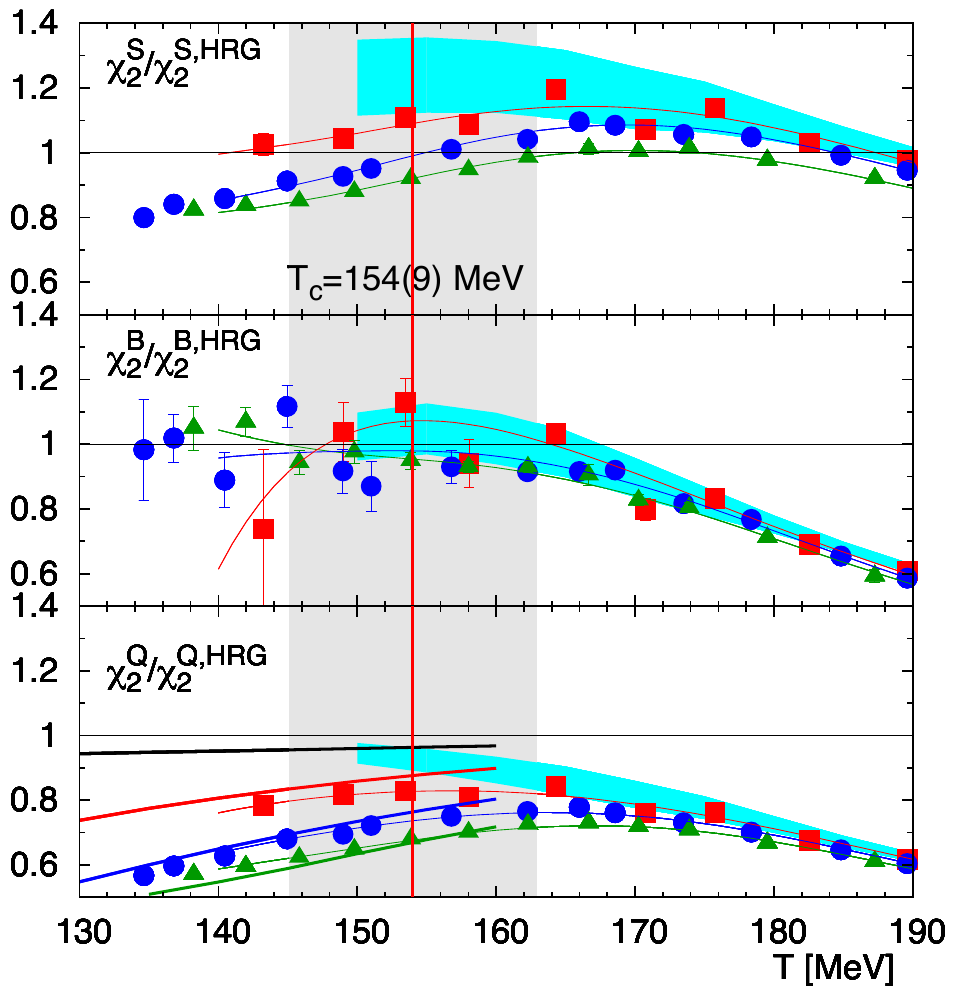}
\end{center}
\end{minipage}
\end{center}
\caption{Cumulants of the net baryon number fluctuations (left) and electric charge fluctuations (middle) up to the sixth order; quadratic cumulant of strangeness, baryon number and electric charge normalized by the HRG results (right). Different symbols denote different lattice spacings: $N_\tau=6$ (triangles), $8$ (circles) and $12$ (squares). Solid lines on the right panel indicate spline fits to the data. For the electric charge fluctuations additional solid lines indicate HRG results with a modified pion mass (see text for explanation). The vertical bar indicates the QCD transition temperature as obtained in \cite{Tc}. \label{fig:chi}}
\label{fig:harmony}
\end{figure}

On the lattice we study derivatives of the partition function with respect to baryon ($B$), electric charge ($Q$) and 
strangeness ($S$) chemical potentials, which are also kwon as generalized susceptibilities and are defined as 
\begin{equation}
\left(VT^3\right)\cdot\chi^{BQS}_{ijk}(T)= \left(\partial^{i+j+k}\ln Z(T,\mu_B,\mu_Q,\mu_S)\right) \left/ \left(\partial \hat\mu_B^i \partial \hat\mu_Q^j \partial \hat\mu_S^k \right)\right.,
\end{equation} 
with $\hat\mu_X=\mu_X/T$ and $X=B,Q,S$. The lattice studies are performed at $\mu_X=0$ and sufficiently close to the thermodynamic limit. The generalized susceptibilities are thus intensive quantities. They can also be interpreted as Taylor expansion coefficients of $\ln Z$ and, furthermore, be related to the cumulants of the fluctuations of the net charges ($N_X$), which are measured in heavy ion collisions. {\it E.g.}, for the diagonal fluctuations one obtains
\begin{eqnarray}
\left(VT^3\right)\cdot\chi^X_2 &=& \left<\left(\delta N_X\right)^2\right>,\\
\left(VT^3\right)\cdot\chi^X_4 &=& \left<\left(\delta N_X\right)^4\right>-3\left<\left(\delta N_X\right)^2\right>^2,\\
\left(VT^3\right)\cdot\chi^X_6 &=& \left<\left(\delta N_X\right)^6\right>-15\left<\left(\delta N_X\right)^4\right>\left<\left(\delta N_X\right)^2\right>+30\left<\left(\delta N_X\right)^2\right>^3,
\end{eqnarray}
with $\delta N_X=N_X-\left<N_X\right>$. In Fig.~\ref{fig:chi} (left and middle) we show our recent results for the 
diagonal fluctuations of net baryon number and electric charge, obtained with an highly improved staggered quark (HISQ) action on lattices with temporal extent of $N_\tau=6$ and $8$. 
In oder to compare the lattice QCD results with experimentally measured fluctuations one has to eliminate the unknown fireball volume by taking ratios of cumulants. Our approach to fix the strangeness and electric charge chemical potentials and to extract the remaining freeze-out parameters, {\it i.e.}  the freeze-out temperature ($T^f$) and the freeze-out chemical potential ($\mu_B^f$), from a comparison of lattice and experimental results of ratios of cumulants of the electric charge fluctuations has recently been proposed by us \cite{PRL} and was also presented at this conference \cite{Swagato}. At nonzero baryon density the method is based on a next to leading order (NLO) Taylor expansion of two different ratios. In general we find that the NLO contributions are below 10\% in the $\hat \mu_B$ range relevant for the RHIC energy scan down to collision energies of $\sqrt{s}\gtrsim 20$~GeV.

\section{O(4) critical behavior}
In the chiral limit of QCD with two massless quarks the transition from the hadronic world 
to the quark gluon plasma is presumably of second order (see Fig.~\ref{fig:pdiag}). 
At vanishing chemical potential the critical point is expected to be in the universality class of the three dimensional $O(4)$ symmetric spin model 
and is expected to persist even in the presence of a physical strange quark mass. It is illuminating to perform a universal scaling 
analysis that is connected with the critical point in the chiral limit. Here to leading order the chemical potential 
enters only in the temperature like scaling field
\begin{equation}
t=t_0^{-1}\left((T-T_0)/T_0 + \kappa_B \hat\mu_B^2+ \kappa_S \hat\mu_S^2 + \kappa_{BS}\hat\mu_B \hat\mu_S  \right),
\label{eq:t}
\end{equation} 
as a finite chemical potential does not alter chiral symmetry breaking. It is thus easy to see that at $\mu_X=0$ the contribution from the singular part of the partition functions to the generalized susceptibilities follows the pattern 
\cite{Ejiri:2005wq}, 
\begin{equation}
\chi^X_{(2n)}\sim |t|^{2-n-\alpha},
\end{equation}
where $\alpha$ is the critical exponent of the specific heat, which is small and negative. The fourth order cumulants will thus develop a cusp in the chiral limit, whereas the sixth order cumulants are divergent with amplitudes that have different signs below and above $T_c$. The lattice data is qualitatively consistent with that picture as can be seen in Fig.~\ref{fig:chi} (left and middle). A more detailed scaling analysis using a recent parametrization of the $O(4)$-scaling function of the specific heat \cite{O4} is work in progress. It will allow to obtain various non-universal normalization constants that map QCD to the universal $O(4)$-symmetric theory such as $T_0,t_0,h_0,\kappa_X,\kappa_{XY}$. Some of them are of immediate interest, such as the transition temperature in the chiral limit $T_0$ or the curvature $\kappa_X$ that characterizes the change of the transition temperature in the direction of the chemical potential $\mu_X$. Recently the HotQCD Collaboration and the BNL-Bielefeld Collaboration performed a scaling analysis of the chiral condensate, the chiral susceptibility \cite{magnetic, Tc} and a particular mixed susceptibility \cite{kappa}.  That also determined the normalization constant $z_0=h_0^{1/\beta\delta}/t_0$, where $\beta$ and $\delta$ are critical exponents, which fixes the quark mass dependence of the QCD transition (at $\mu_X=0)$. This was used by HotQCD in oder to determine the crossover temperature $T_c=154(9)$~MeV \cite{Tc} at physical quark masses.

\section{Comparison with the Hadron Resonance Gas}
Below the QCD transition temperature we can compare our lattice results with the statistical Hadron Resonance Gas (HRG) model, which describes the hadronization process in heavy ion collisions quite successfully \cite{HRG}. For the second order cumulants of baryon number, electric charge and strangeness fluctuations this comparison is performed in Fig.~\ref{fig:chi} (right) \cite{c2}. Here we plot the lattice data normalized by the corresponding HRG results. Light blue bands indicate the continuum extrapolations based on lattices with temporal extent $N_\tau=6,8,12$. We find that the continuum extrapolations of $\chi^B_2$ and $\chi^Q_2$ approach the HRG from below and are in agreement with the HRG at temperatures up to $T\sim (150-160)$ MeV. The strangeness fluctuations, however, seem to overshoot the HRG and eventually approach the HRG at much lower temperatures. Note that for the electric charge fluctuations additional solid lines at low temperature indicate HRG results with a modified pion mass. We have chosen the pion masses such that they agree with the averaged pion mass (root-mean-square mass $m_\pi^{\rm RMS}$) of the pion spectrum on the lattice at given $N_\tau$. At finite $N_\tau$ the pion spectrum of staggered fermions on the lattice achieve an unphysical splitting which leads to an increased $m_\pi^{\rm RMS}$. In general, we find that the modified HRG provides a good approximation to the electric charge fluctuations below $T_c$. 

A comparison of the HRG with higher order cumulants is at present not very meaningful as we do not have a continuum result yet.  
Moreover, for the electric charge fluctuations, which can in principle be immediately compared to the experimental results, the distorted pion spectrum on the lattice becomes increasingly problematic, as the higher order cumulants are increasingly sensitive to an increased $m_\pi^{\rm RMS}$. One thus has to use even finer lattice spacings such as $N_\tau=16$ in order to control this systematic effect. In fact, we find that the HRG results for $\chi^Q_4$ with a pion mass corresponding to the $m_\pi^{\rm RMS}$ of our $N_\tau=6$ and $8$ lattices are indistinguishable from that of a HRG with an infinitely heavy pion mass.

\end{document}